\documentclass[journal]{IEEEtran}

\ifCLASSINFOpdf
\else
   \usepackage[dvips]{graphicx}
\fi
\usepackage{url}

\usepackage{graphicx}
\usepackage{amsmath,bm,graphicx}
\usepackage{amsfonts}
\usepackage{multirow,makecell}
\usepackage{algorithmic,algorithm}
\usepackage{hyperref}
\usepackage{enumitem}
\usepackage{adjustbox} 
\usepackage{bm}  
\usepackage{hhline}
\usepackage{booktabs}
\usepackage{multirow}
\usepackage{adjustbox}
\usepackage{pifont} 
\usepackage{xcolor}
\usepackage{url}
\usepackage{color}
\usepackage{graphicx}
\usepackage{amsmath,bm,graphicx}
\usepackage{amsfonts}
\usepackage{multirow,makecell}
\usepackage{algorithmic,algorithm}
\usepackage{hyperref}
\usepackage{enumitem}
\usepackage{adjustbox} 
\usepackage{bm}  
\usepackage{pifont}
\colorlet{blue}{black}
\colorlet{red}{black}
\usepackage[most]{tcolorbox}
\begin{document}

\title{UMelCodec: An Ultra-Low-Bitrate and Lightweight Mel-Domain Neural Speech Codec}

\author{Hui-Peng Du, Yang~Ai,~\IEEEmembership{Member,~IEEE},~Zhen-Hua~Ling,~\IEEEmembership{Senior Member,~IEEE}
\thanks{This work was funded by the National Nature Science Foundation of China under Grant 62301521. (Corresponding author: Yang~Ai)}
\thanks{Hui-Peng Du, Yang Ai, and Zhen-Hua Ling are with the National Engineering Research Center of Speech and Language Information Processing, University of Science and Technology of China, Hefei, 230027, China (e-mail: redmist@mail.ustc.edu.cn, yangai@ustc.edu.cn, zhling@ustc.edu.cn).}}

\maketitle

\begin{abstract}
Ultra-low-bitrate speech coding is crucial for bandwidth-constrained communication and compact storage, but preserving intelligibility, naturalness, and fine-grained acoustic details at bitrates of only a few hundred bps remains challenging.
Existing purely acoustic codecs generally operate at relatively high bitrates, whereas incorporating semantic representations can reduce the bitrate, but at the cost of increased complexity.
Therefore, we propose UMelCodec, a purely acoustic neural speech codec operating in the mel domain at an ultra-low bitrate of 325~bps with a lightweight architecture.
UMelCodec consists of two stages. 
In the first mel-domain coding stage, alias-free Snake-ConvNeXt (AF-SC) encoder–decoder backbones model temporal and harmonic structures, a factorized single-codebook vector quantizer (VQ) enables ultra-low-bitrate quantization, and a multi-scale mel-spectrogram discriminator (MSMD) improves mel-spectrogram reconstruction. 
In the second waveform synthesis stage, a lightweight modified discrete cosine transform (MDCT)-based neural vocoder predicts the MDCT spectrum from the reconstructed mel spectrogram and recovers the waveform through inverse MDCT.
Experiments on LibriTTS dataset demonstrate that UMelCodec outperforms strong baselines across most objective reconstruction metrics while maintaining competitive subjective naturalness. In terms of model complexity and inference efficiency, UMelCodec has only 19.67~M parameters, requires \textcolor{blue}{1.93}~G MACs, and supports real-time inference on both GPU and CPU platforms.
\end{abstract}

\IEEEpeerreviewmaketitle
\vspace{-3mm}
\section{Introduction}
\label{sec:introduction}

\IEEEPARstart{S}{peech} coding aims to represent speech signals with compact bitstreams and reconstruct intelligible and natural waveforms at decoder side. 
It is a key technique for bandwidth-constrained communication, speech storage, etc.

Traditional speech codecs rely on carefully designed signal-processing modules. Opus~\cite{valin2012opus} and EVS~\cite{dietz2015overview} provide robust speech quality but generally operate at bitrates of several kbps or higher. 
Recent acoustic-level neural codecs improve rate--distortion performance through learned signal representations. 
Among them, waveform-domain codecs, e.g., SoundStream~\cite{zeghidour2021soundstream}, EnCodec~\cite{defossez2023encodec}, DAC~\cite{kumar2023dac}, and BigCodec~\cite{xin2024bigcodec}, directly model time-domain signals using encoder--quantizer--decoder architectures, but their deeply stacked downsampling and upsampling layers incur substantial computational cost. 
Spectral-domain codecs reduce the modeling burden by exploiting structured time--frequency representations: APCodec~\cite{ai2024apcodec}, MDCTCodec~\cite{jiang2024mdctcodec}, and Spectral Codec~\cite{langman2024spectral} quantize amplitude--phase spectral, modified discrete cosine transform (MDCT) spectral, and mel-spectral representations, respectively.
However, these codecs still operate at relatively high bitrates, and preserving fine-grained acoustic details in the few-hundred-bps regime remains challenging. To address this bitrate limitation at the acoustic level,
\textcolor{blue}{WavTokenizer~\cite{ji2025wavtokenizer},
Mimi~\cite{defossez2024moshi}, and Stable Codec~\cite{parker2025scaling}
achieve low-bitrate coding with relatively complex architectures, while
SNAC~\cite{siuzdak2024snac} improves compression efficiency. Nevertheless, few-hundred-bps coding with low
computational complexity remains challenging.}
\textcolor{blue}{FlowMAC~\cite{pia2025flowmac},  MuCodec~\cite{xu2024mucodec}, and FMelCodec~\cite{du2026fmelcodec} introduce} flow-matching-based refinement~\cite{lipman2023flow} for \textcolor{blue}{ultra-low-bitrate} mel-spectrogram coding, but its iterative mel refinement and waveform generation require multiple inference steps and increase computational complexity.
Alternatively, semantic-assisted codecs, e.g., SemantiCodec~\cite{liu2024semanticodec} and FocalCodec~\cite{dellalibera2025focalcodec}, achieve \textcolor{blue}{ultra-low-bitrate} coding by leveraging high-level representations from large pretrained models; however, the additional upstream models substantially increase the overall model size and computational cost and may discard fine-grained acoustic information. 
Therefore, jointly achieving an ultra-low bitrate, low computational complexity, and high acoustic fidelity remains a challenge. 


To address these limitations, we develop \textbf{UMelCodec}, an \textbf{U}ltra-low-bitrate and lightweight \textbf{Mel}-domain neural speech \textbf{Codec} that relies solely on acoustic representations and operates at only 325~bps.
\textcolor{red}{The key contribution lies in jointly designing harmonic-aware mel-domain compression and efficient waveform synthesis to balance reconstruction quality, bitrate and complexity.
Specifically, }the mel-domain coding stage uses alias-free Snake-ConvNeXt (AF-SC) encoder--decoder backbones, a factorized single-codebook vector quantizer (VQ), and a multi-scale mel-spectrogram discriminator (MSMD) to compress and reconstruct mel spectrograms at an ultra low bitrate. 
The waveform synthesis stage then employs a lightweight vocoder to predict MDCT spectra from the reconstructed mel spectrograms and recover the waveforms. Experiments show that UMelCodec surpasses strong baselines on most objective reconstruction metrics while achieving competitive subjective naturalness. 
Moreover, UMelCodec maintains low model complexity and computational cost while supporting real-time inference across different hardware platforms.

\vspace{-2mm}
\section{Proposed Method}
\label{sec:method}

\subsection{Overview}
\label{ssec:overview}

\begin{figure*}[t]
    \centering
    \includegraphics[width=0.97\linewidth]{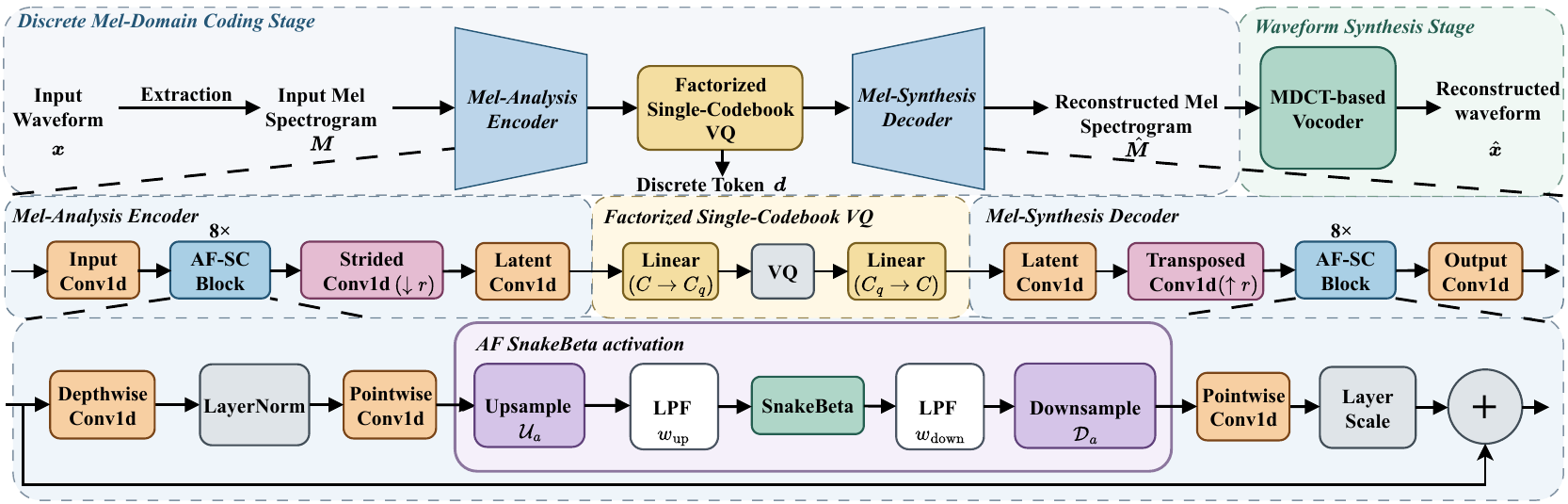}
    \caption{\textcolor{blue}{Overall framework of the proposed UMelCodec.}}
    \label{fig:framework}
\end{figure*}
As illustrated in Fig.~\ref{fig:framework}(a), the proposed UMelCodec is a two-stage ultra-low-bitrate speech coding framework. 
Let $\bm{x}\in\mathbb{R}^{T}$ and $\hat{\bm{x}}\in\mathbb{R}^{T}$ denote the input and reconstructed waveforms sampled at $f_s$, respectively, where $T$ is the number of waveform samples. 
In the first mel-domain coding stage, a mel spectrogram $\bm{M}\in\mathbb{R}^{N\times D}$ is first extracted from $\bm{x}$ using a hop size of $H$ samples, yielding $N=T/H$ mel frames, where $D$ is the number of mel bins. 
The mel-domain codec then maps $\bm{M}$ to a discrete index sequence $\bm{d}=[d_1,d_2,\ldots,d_{N'}]^\top\in\mathbb{N}^{N'}$ and reconstructs mel-spectrogram $\hat{\bm{M}}\in\mathbb{R}^{N\times D}$, where each $d_n\in\{1,\ldots,K\}$, $K$ is the codebook size, and $N'$ is the number of latent frames after downsampling. 
In the second waveform synthesis stage, an MDCT-based vocoder reconstructs $\hat{\bm{x}}$ from $\hat{\bm{M}}$.
\vspace{-2mm}
\vspace{-1mm}
\subsection{Mel-Domain Coding Stage}

As shown in Fig. \ref{fig:framework}(b), the mel codec consists of a mel-analysis encoder, a factorized single-codebook VQ, and a mel-synthesis decoder. Given an input mel spectrogram $\bm{M}$, the mel-analysis encoder first applies an input convolution to map each $D$-dimensional mel frame into a higher-dimensional feature space. The resulting sequence is processed by eight AF-SC blocks to model temporal dependencies and periodic acoustic structures. A strided convolution then reduces the temporal frame rate by an overall factor of $r$, yielding $N'=N/r$ latent frames. Finally, a latent convolution reduces the feature dimension and produces the latent sequence $\bm{Z}\in\mathbb{R}^{N'\times C}$ for quantization, where $C$ denotes the latent feature dimension.

As shown in Fig.~\ref{fig:framework}(c), AF-SC block follows a ConvNeXt-style architecture~\cite{liu2022convnext,woo2023convnextv2}. 
A large-kernel depthwise convolution first models temporal context, followed by layer normalization and a pointwise projection that expands the feature dimension. 
An alias-free (AF) SnakeBeta activation \cite{ziyin2020snake,lee2023bigvgan} introduces a periodic inductive bias for modeling harmonic structures, after which a second pointwise projection restores the input dimension. Layer scaling and a residual connection are further applied to stabilize optimization and preserve input information. AF-SnakeBeta applies the SnakeBeta nonlinearity at an increased temporal resolution, with fixed low-pass filters (LPFs) placed before and after the activation:
\begin{equation}
    \mathrm{AF\text{-}SnakeBeta}(\bm{f})
    =
    \mathcal{D}_{a}\!\left(
    \bm{w}_{\mathrm{down}} *
    \mathrm{SnakeBeta}\!\left(
    \bm{w}_{\mathrm{up}} *
    \mathcal{U}_{a}(\bm{f})
    \right)\right),
\end{equation}
where $\bm{f}$ denotes the input feature sequence, $\mathcal{U}_{a}(\cdot)$ and $\mathcal{D}_{a}(\cdot)$ denote upsampling and downsampling by a factor of $a$, respectively, and $*$ denotes convolution. The fixed Kaiser-windowed sinc LPF kernels $\bm{w}_{\mathrm{up}}$ and $\bm{w}_{\mathrm{down}}$ suppress the spectral images introduced by upsampling and remove the high-frequency components generated by SnakeBeta before downsampling, respectively, thereby reducing aliasing.

The latent sequence $\bm{Z}$ produced by the mel-analysis encoder is subsequently discretized using the factorized single-codebook VQ. Following~\cite{kumar2023dac}, each $C$-dimensional latent vector is projected to a lower-dimensional quantization space of dimension $C_q$, where $C_q<C$, and matched to its nearest entry in a codebook of $K$ vectors. The resulting frame-level indices form the discrete sequence $\bm{d}$. The selected codebook vectors are then projected back to dimension $C$ before being passed to the mel-synthesis decoder. This $C\!\rightarrow\!C_q\!\rightarrow\!C$ factorization facilitates code assignment under the single-codebook setting. 
The resulting theoretical bitrate is $R=\frac{f_s}{Hr}\log_2 K$ (bps).

The mel-synthesis decoder reverses the transformations of the mel-analysis encoder. A latent convolution first expands the quantized representation to the decoder feature dimension. A transposed convolution then upsamples the temporal sequence by a factor of $r$, restoring the original mel frame rate and sequence length $N$. The upsampled features are further refined by eight AF-SC blocks, and an output convolution maps them back to $D$ mel bins to obtain the reconstructed mel spectrogram $\hat{\bm{M}}$.

\vspace{-4mm}
\subsection{Waveform Synthesis Stage}
\label{ssec:vocoder}

The neural waveform synthesis stage reconstructs the time-domain waveform from the decoded mel spectrogram using a lightweight MDCT-based vocoder. 
First, an amplitude-spectrum estimate is derived from $\hat{\bm{M}}$ using the pseudo-inverse of the mel filter-bank matrix. The estimated spectrum is then processed by an input projection and eight ConvNeXt v2 blocks \cite{woo2023convnextv2} to model temporal and spectral dependencies, followed by an output projection that predicts the corresponding MDCT spectrum. Finally, the reconstructed waveform $\hat{\bm{x}}$ is obtained through inverse MDCT.

\vspace{-4mm}
\subsection{Training Strategy}
\vspace{-1.5mm}
\label{ssec:training_strategy}

The modules in mel-domain coding stage and neural waveform synthesis stage are trained separately. 
For the mel-domain codec, the training objective comprises reconstruction, quantization, and adversarial losses. The mel-spectrogram reconstruction loss $\mathcal{L}_{\mathrm{rec}}$ measures the equal-weight $\ell_1$ and $\ell_2$ distances between $\bm{M}$ and $\hat{\bm{M}}$. Following VQ-VAE~\cite{oord2017vqvae}, the quantization objective includes a commitment loss $\mathcal{L}_{\mathrm{com}}$ and a codebook loss $\mathcal{L}_{\mathrm{cb}}$. To improve the perceptual and structural fidelity of the reconstructed mel spectrograms, we employ an MSMD comprising three sub-discriminators at progressively downsampled temporal resolutions to distinguish real and reconstructed mel spectrograms. The resulting adversarial loss $\mathcal{L}_{\mathrm{adv}}$ and feature-matching loss $\mathcal{L}_{\mathrm{fm}}$ encourage realistic mel-spectral patterns and stabilize adversarial training, respectively. The overall objective is
\vspace{-1mm}
\begin{equation}
\begin{aligned}
    \mathcal{L}_{G}
    =
    &\lambda_{\mathrm{rec}}\mathcal{L}_{\mathrm{rec}}
    +\lambda_{\mathrm{com}}\mathcal{L}_{\mathrm{com}}
    +\lambda_{\mathrm{cb}}\mathcal{L}_{\mathrm{cb}} \\
    &+\lambda_{\mathrm{adv}}\mathcal{L}_{\mathrm{adv}}
    +\lambda_{\mathrm{fm}}\mathcal{L}_{\mathrm{fm}},
\end{aligned}
\end{equation}
where $\lambda_{\mathrm{rec}}$, $\lambda_{\mathrm{com}}$, $\lambda_{\mathrm{cb}}$, $\lambda_{\mathrm{adv}}$, and $\lambda_{\mathrm{fm}}$ are hyperparameters.

The MDCT-based vocoder is trained using an MDCT-domain multi-scale discriminator, which processes MDCT spectra at three resolutions through parallel two-dimensional convolutional branches. Following the multi-scale adversarial learning strategy~\cite{kumar2019melgan,kong2020hifigan}, its objective combines adversarial and feature-matching losses with MDCT-spectrum and mel-spectrogram reconstruction losses.

\vspace{-3mm}
\section{Experimental Setups}
\vspace{-2mm}
\label{sec:pagestyle}
\subsection{Dataset}
We conducted experiments on the LibriTTS corpus~\cite{zen2019libritts}, with all speech clips downsampled to $f_s=$16 kHz. 
For training, we used the full official training set, including \textit{train-clean-100}, \textit{train-clean-360}, and \textit{train-other-500}. The UMelCodec was evaluated on \textit{test-clean} and \textit{test-other} to assess reconstruction quality under clean and more challenging acoustic conditions.

\vspace{-4mm}
\subsection{Implementation}
\label{ssec:implementation}
\vspace{-1mm}
For UMelCodec\footnote{Audio samples can be found at \url{https://redmist328.github.io/UMelCodecdemo/}.}, mel spectrograms were extracted from waveforms using an FFT size of 1024, a window length of 640, a hop size of $H=160$, and $D=80$ mel bins. 
For the AF-SC blocks in both the mel-analysis encoder and mel-synthesis decoder, the feature and intermediate dimensions were set to 256 and 512, respectively, while the resampling factor $a$ in the AF-SnakeBeta activation was set to 2.
The temporal downsampling/upsampling factor was set to $r=4$. 
The latent dimension, quantization dimension, and codebook size were set to $C=32$, $C_q=8$, and $K=8192$, respectively, corresponding to a theoretical bitrate of 325~bps. The loss weights $\lambda_{\mathrm{rec}}$, $\lambda_{\mathrm{com}}$, $\lambda_{\mathrm{cb}}$, $\lambda_{\mathrm{adv}}$, and $\lambda_{\mathrm{fm}}$ were set to 45, 2.5, 10, 0.1, and 0.1, respectively. AdamW~\cite{loshchilov2019decoupled} optimizer was used for training with a learning rate of $2\times10^{-4}$, $\beta_1=0.8$, $\beta_2=0.99$, and an exponential decay factor of 0.999. 
The separately trained MDCT-based vocoder employed ConvNeXt v2 blocks, each with feature and intermediate dimensions of 513 and 1536, respectively.
The MDCT spectra were computed 
\textcolor{blue}{using a 320-sample cosine window with a hop size of 160, yielding 160 frequency bins.}
\vspace{-4mm}
\subsection{Baselines}
\label{ssec:baselines}

We compared UMelCodec with DAC~\cite{kumar2023dac}, BigCodec~\cite{xin2024bigcodec}, FMelCodec~\cite{du2026fmelcodec}, SemantiCodec~\cite{liu2024semanticodec}, and FocalCodec~\cite{dellalibera2025focalcodec}. 
Since DAC, BigCodec, and FMelCodec do not provide official 325-bps checkpoints, we retrained them on the same dataset using a 25-Hz frame rate and a single 8192-entry codebook. For SemantiCodec and FocalCodec, we used their official 325-bps checkpoints.
\vspace{-3mm}
\subsection{Evaluation Metrics}
\vspace{-2mm}
\label{ssec:metrics}
We evaluated speech reconstruction quality using both objective and subjective metrics. 
The objective metrics included the virtual speech quality objective listener (ViSQOL)~\cite{hines2015visqol}, log-spectral distance (LSD), mel-cepstral distortion (MCD) \cite{kubichek1993mel}, and short-time objective intelligibility (STOI) \cite{taal2011algorithm}. 
For subjective evaluation, mean opinion score (MOS) tests were conducted on Amazon Mechanical Turk platform, where each batch contained 20 utterances rated by at least 25 native English listeners on a five-point naturalness scale with 0.5-point increments. Model complexity and inference efficiency were evaluated on the LibriTTS \textit{test-other} set using the number of parameters (Param.), multiply--accumulate operations (MACs), and real-time factor (RTF). MACs were calculated using a \textcolor{blue}{1}-second speech input, while GPU and CPU real-time factors (RTFs) were measured on an NVIDIA GeForce RTX 3090 and an Intel Xeon E5-2680 v3 processor, respectively.

\begin{table*}[t]
    \centering
    \caption{Objective and subjective evaluation results of UMelCodec and baseline codecs at 325~bps on the LibriTTS test-clean and test-other sets. Here, $b\times$ denotes an inference speed of $b$ times real time. The best and second-best results are highlighted in bold and underlined, respectively. MOS values are reported with 95\% confidence intervals.}
    \label{tab:main_results}
    \setlength{\tabcolsep}{2.2pt}
    \renewcommand{\arraystretch}{1.02}
    \scriptsize
    \adjustbox{width=0.9\textwidth}{
    \begin{tabular}{l|ccccc|ccccc|cccc}
        \hline
        \multirow{4}{*}{Model}
        & \multicolumn{10}{c|}{\textbf{\textcolor{red}{Quality Metrics}}}
        & \multicolumn{4}{c}{\textbf{\textcolor{red}{Complexity Metrics}}}\\
        \cline{2-15}
        & \multicolumn{5}{c|}{\textit{test-clean}}
        & \multicolumn{5}{c|}{\textit{test-other}}
        & \multirow{3}{*}{\shortstack{Param.\\(M)$\downarrow$}}
        & \multirow{3}{*}{\shortstack{MACs\\(G)$\downarrow$}}
        & \multirow{3}{*}{\shortstack{GPU RTF\\$\downarrow$}}
        & \multirow{3}{*}{\shortstack{CPU RTF\\$\downarrow$}}\\
        \cline{2-11}
        & ViSQOL & LSD & MCD & STOI & MOS
        & ViSQOL & LSD & MCD & STOI & MOS
        & & & & \\[-1mm]
        & $\uparrow$ & (dB)$\downarrow$ & (dB)$\downarrow$
        & $\uparrow$ & $\uparrow$
        & $\uparrow$ & (dB)$\downarrow$ & (dB)$\downarrow$
        & $\uparrow$ & $\uparrow$
        & & & & \\
        \hline

        DAC
        &3.17&1.04&4.10&0.79&3.40 (\ensuremath{\pm}0.08)
        &2.91&1.19&4.67&0.75&2.86 (\ensuremath{\pm}0.08)
        &76.71&\textcolor{blue}{47.80}
        &0.022 (45.58\ensuremath{\times})
        &1.546 (0.65\ensuremath{\times})\\

        BigCodec
        &3.38&\underline{1.00}&\underline{3.47}&\underline{0.83}
        &3.53 (\ensuremath{\pm}0.07)
        &3.14&1.16&\underline{3.96}&\underline{0.80}
        &3.21 (\ensuremath{\pm}0.08)
        &114.76&\textcolor{blue}{44.62}
        &0.030 (32.88\ensuremath{\times})
        &4.311 (0.23\ensuremath{\times})\\

        FMelCodec
        &\underline{3.56}&1.02&4.11&0.81
        &3.33 (\ensuremath{\pm}0.08)
        &\underline{3.29}&\textbf{1.06}&4.42&0.77
        &2.51 (\ensuremath{\pm}0.08)
        &\underline{30.62}&\textcolor{blue}{18.76}
        &0.033 (30.75\ensuremath{\times})
        &0.713 (1.40\ensuremath{\times})\\

        SemantiCodec
        &3.35&1.18&4.20&0.78&3.10 (\ensuremath{\pm}0.08)
        &3.27&1.14&4.39&0.75&2.79 (\ensuremath{\pm}0.08)
        &1877.67&\textcolor{blue}{1603.71}
        &1.685 (0.59\ensuremath{\times})
        &35.940 (0.03\ensuremath{\times})\\

        FocalCodec
        &3.48&1.18&3.70&\underline{0.83}
        &\textbf{3.63} (\ensuremath{\pm}0.08)
        &3.25&1.35&4.55&\underline{0.80}
        &\textbf{3.64} (\ensuremath{\pm}0.07)
        &144.22&\underline{\textcolor{blue}{8.51}}
        &\underline{0.011} (91.16\ensuremath{\times})
        &\underline{0.355} (2.82\ensuremath{\times})\\

        \hline

        UMelCodec
        &\textbf{3.80}&\textbf{0.96}&\textbf{3.13}&\textbf{0.86}
        &\underline{3.59} (\ensuremath{\pm}0.08)
        &\textbf{3.56}&\textbf{1.06}&\textbf{3.85}&\textbf{0.81}
        &\underline{3.30} (\ensuremath{\pm}0.08)
        &\textbf{19.67}&\textbf{\textcolor{blue}{1.93}}
        &\textbf{0.006 (156.37\ensuremath{\times})}
        &\textbf{0.120 (8.36\ensuremath{\times})}\\

        \hline
    \end{tabular}}
\end{table*}

\begin{table}[t]
    \centering
    \caption{Ablation results of UMelCodec on the LibriTTS test-clean set.}
    \label{tab:ablation}
    \setlength{\tabcolsep}{2.2pt}
    \renewcommand{\arraystretch}{1.0}
    \small
    \adjustbox{width=0.95\columnwidth}{
    \begin{tabular}{l|cccc}
        \hhline{-|----}
        Model&ViSQOL&LSD&MCD&STOI\\
        &$\uparrow$&(dB)$\downarrow$&(dB)$\downarrow$&$\uparrow$\\
        \hhline{-|----}
        UMelCodec w/o MSMD
        &3.75&1.00&3.33&\textbf{0.86}\\
        
        UMelCodec w/o AF filtering
        &3.76&0.97&3.16&0.85\\
        
        UMelCodec w/o VQ factorization
        &3.68&0.99&3.49&0.84\\
        \hhline{-|----}
        
        UMelCodec
        &\textbf{3.80}&\textbf{0.96}&\textbf{3.13}&\textbf{0.86}\\
        \hhline{-|----}
    \end{tabular}}
\end{table}
\vspace{-2.5mm}
\section{Experimental Results}
\label{sec:results}
\subsection{Comparison with Baseline Codecs}

Table~\ref{tab:main_results} compared UMelCodec with representative acoustic-level codecs, including waveform- and mel-domain methods, as well as semantic-assisted codecs, all operating at 325~bps. 
UMelCodec achieved the best ViSQOL, MCD, and STOI results on both test sets, together with the best LSD on \textit{test-clean} and the joint-best LSD on \textit{test-other}. 
Compared with the waveform-domain DAC and BigCodec, UMelCodec provided higher perceptual quality, intelligibility, and spectral fidelity, demonstrating the effectiveness of speech coding based on compact mel-spectral representations. 
It also outperformed the generative FMelCodec without requiring iterative flow-matching refinement. 
As a purely acoustic codec, UMelCodec achieved higher full-reference reconstruction fidelity than SemantiCodec and FocalCodec without relying on large pretrained representation models, suggesting that their semantic representations may prioritize linguistic content over fine-grained acoustic details.

The subjective results further supported the objective findings to some extent. 
On \textit{test-clean}, UMelCodec achieved 
\textcolor{blue}{a MOS of 3.59$\pm$0.08,
close to the 3.63$\pm$0.08 obtained by FocalCodec, while their
95\% confidence intervals overlap.}
On \textit{test-other}, FocalCodec obtained a higher MOS despite its lower objective performance. 
This may be because its pretrained semantic representations suppress part of the background interference and generate cleaner speech rather than strictly preserving the original acoustic environment. 
In contrast, UMelCodec maintained competitive subjective naturalness while more faithfully preserving the input acoustic characteristics.

In terms of model complexity and inference efficiency, UMelCodec required only 19.67~M parameters and \textcolor{blue}{1.93}~G MACs, both the lowest among the evaluated codecs. 
According to the RTF measurements, it achieved inference speeds of 156.37 times real time on GPU and 8.36 times real time on CPU, demonstrating efficient deployment on both platforms.
\vspace{-5mm}
\subsection{Ablation Studies}

We then conducted ablation experiments to investigate the contributions of the main components in UMelCodec, as reported in Table~\ref{tab:ablation}. 
{UMelCodec w/o MSMD} exhibited lower ViSQOL and higher LSD and MCD than the full model, while achieving the same STOI. This result suggests that multi-scale adversarial supervision \textcolor{blue}{contributes to} the perceptual quality and spectral fidelity of the reconstructed mel spectrograms by encouraging realistic temporal structures at different resolutions. 
{UMelCodec w/o AF filtering}, which removed the upsampling, downsampling, and LPF operations from the AF-SnakeBeta activation, performed \textcolor{blue}{slightly} worse than the full model across all four objective metrics. This \textcolor{blue}{suggests} that suppressing aliasing around the periodic nonlinearity benefited perceptual quality, spectral reconstruction, and speech intelligibility \textcolor{blue}{to some extent}. 
{UMelCodec w/o VQ factorization} also exhibited consistent degradation across all metrics, particularly in ViSQOL, MCD, and STOI, indicating that low-dimensional code assignment facilitated effective quantization and preserved perceptually important information under the single-codebook setting.
\vspace{-1mm}
\vspace{-2mm}
\subsection{Analysis of the Effect of Codebook Size}
\label{ssec:codebook_size}
\begin{figure}[t]
    \centering
    \includegraphics[width=0.9\linewidth]{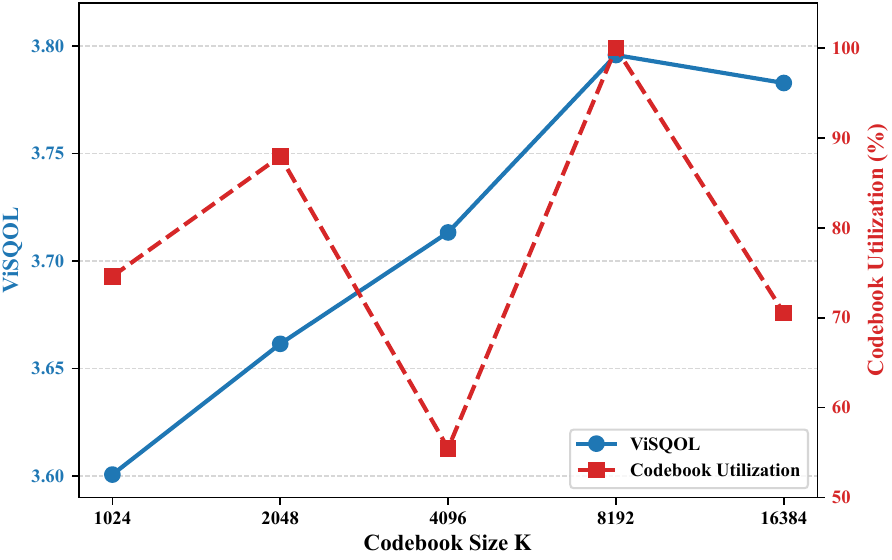}
    \caption{\textcolor{blue}{Effect of the codebook size $K$ on ViSQOL and codebook utilization on the LibriTTS test-clean set.}}
    \label{fig:codebook_size}
\end{figure}

The codebook size determined both the representation capacity and the utilization efficiency of the factorized single-codebook VQ\textcolor{blue}{, which is the percentage of codebook selected at least once over the
entire test set.}
We therefore analyzed its effect to justify the choice of 8192 entries and examine the trade-off between reconstruction quality and codebook utilization. 
As shown in Fig.~\ref{fig:codebook_size}, ViSQOL generally improved as the codebook size increased and reached its maximum at 8192, where all codebook entries were utilized. 
Further increasing the codebook size to 16384 reduced the utilization to approximately 70\% and slightly degraded ViSQOL. 
In this case, although the nominal codebook size was doubled, the number of utilized entries increased by only about 40\% compared with the fully utilized 8192-entry codebook, indicating that many additional entries were redundant or rarely selected. 
Therefore, enlarging the codebook beyond 8192 provided limited effective capacity while making quantization less efficient and degrading reconstruction quality. 
Based on this trade-off, a codebook size of 8192 was adopted in UMelCodec.

\vspace{-2mm}
\section{Conclusion}

This paper presented UMelCodec, an ultra-low-bitrate and lightweight mel-domain neural speech codec operating at 325~bps without relying on auxiliary pretrained semantic models or generative models.
UMelCodec employs AF-SC encoder--decoder backbones, a factorized single-codebook VQ, and an MSMD for mel-spectrogram compression and reconstruction. 
A lightweight neural vocoder predicts the MDCT spectrum from mel-spectrograms and recovers the waveform through inverse MDCT. 
Experiments on LibriTTS show that UMelCodec outperforms strong baselines in objective reconstruction quality while maintaining competitive subjective naturalness. 
In addition, UMelCodec requires only 19.67~M parameters and 1.93~G MACs and supports real-time GPU and CPU inference. 
Ablation studies confirm the effectiveness of MSMD, alias-free activation, and VQ factorization, while codebook analysis favors 8192 entries for balancing reconstruction quality and codebook utilization.
\clearpage

\bibliographystyle{IEEEtran}
\bibliography{mybib}

\end{document}